\documentclass[aps,prb,reprint,longbibliography,superscriptaddress,floatfix]{revtex4-2}

\usepackage{amsmath,amssymb,bm,xcolor,pgfpict2e,graphicx}

\usepackage[hidelinks]{hyperref}

\definecolor{curveblue}{rgb}{0.08,0.30,0.64}

\definecolor{curvered}{rgb}{0.75,0.24,0.10}

\definecolor{curvegreen}{rgb}{0.10,0.48,0.30}

\newcommand{\ii}{\mathrm{i}}

\newcommand{\Tr}{\operatorname{Tr}}

\newcommand{\D}{\mathcal D}

\newcommand{\LL}{\mathcal L}

\newcommand{\PP}{\mathcal P}

\newcommand{\ket}[1]{\lvert#1\rangle}

\newcommand{\bra}[1]{\langle#1\rvert}

\begin{document}

\title{Odd-Even Effects on Liouvillian Spectra and Relaxation Dynamics in Boundary-Driven Spin Chains}

\author{Zhen-Yu Zheng}

\email{zhengzhenyu@sicau.edu.cn}

\affiliation{College of Science, Sichuan Agricultural University, Ya'an 625014, China}

\author{Shu Chen}

\email{schen@iphy.ac.cn}

\affiliation{Beijing National Laboratory for Condensed Matter Physics, Institute of Physics, Chinese Academy of Sciences, Beijing 100190, China}

\affiliation{School of Physical Sciences, University of Chinese Academy of Sciences, Beijing 100049, China}

\date{October 19, 2026}

\begin{abstract}
We investigate odd-even effects on Liouvillian spectra and relaxation dynamics in boundary-driven XY and cluster chains. With the bulk couplings and boundary dissipation rule fixed, the parity of chain length determines whether non-decaying oscillatory modes exist. Consequently, odd- and even-sized chains exhibit qualitatively different dynamical behaviors. We trace this mechanism to the coupling between bulk Majorana paths and the boundary jump operators. An isolated path supports undamped modes, and products of an even number of its Majorana operators span a non-decaying subspace of even operators. This structure yields nonzero purely imaginary eigenvalues of the Liouvillian and persistent contributions to bond correlations or magnetization after preparation in product states. The mechanism extends to cluster interactions of even range under a fixed boundary dissipation rule. The existence and frequencies of the persistent modes are independent of the magnitudes of the positive boundary dissipation rates. Our results unveil how chain parity controls the existence of persistent dynamics.
\end{abstract}

\maketitle

\raggedbottom

\section{Introduction}
Understanding relaxation and information retention in quantum many-body systems is a central problem in nonequilibrium physics~\cite{Eisert2015,Gogolin2016}. Advances in the control of dissipation have stimulated extensive interest in open quantum systems and their collective behavior beyond thermal equilibrium~\cite{Breuer2002,Manzano2020,Fazio2025,Sieberer2025}. A particularly important class is boundary-driven systems, where the environment acts only near the boundaries and the bulk interactions determine how dissipation propagates through the system~\cite{Landi2022}. Such systems exhibit rich stationary and dynamical phenomena, including long-range correlations and nonequilibrium transitions~\cite{ProsenPizorn2008,Yao2025}. Exactly solvable models and analytical approaches have provided valuable insights into their Liouvillian spectra and relaxation dynamics~\cite{Prosen2008,Prosen2010,Barthel2022,Thompson2023,Yamanaka2023,ShibataKitaev2019,ShibataIsing2019,Shibata2020,Zheng2023,Teretenkov2024,WangXL}.

Much of the discussion of open many-body systems has focused on phases in the thermodynamic limit and on the size dependence of relaxation times~\cite{Minganti2018,Znidaric2015,Mori2020,Mori2023,ZhouBZ}. From the dynamical perspective, considerable attention has been paid to the Liouvillian spectrum, the structure of its eigenmodes, and their overlaps with initial states and observables~\cite{Albert2014,Albert2016,Song2019,Haga2021,Yang2022}. Related studies have established routes to information retention through decoherence-free structures and symmetries~\cite{Zanardi1997,Lidar1998,Nigro2019,Buca2012}, spatially restricted dissipation~\cite{Pocklington2022,Andreys2019}, and boundary modes~\cite{Kitaev2001,Diehl2011,Fendley2016,Kemp2017,Kemp2020}. Dissipative spectral structure and selective coupling to the environment further influence the survival of boundary excitations~\cite{Lieu2020,Zheng2025,Yang2023,Wang2026}, while dynamical symmetries can support persistent coherent motion and synchronization~\cite{Buca2019,Tindall2020,Buca2022,Wachtler2024,Matsui2024}. These results also highlight the importance of finite size in interpreting non-decaying dynamics~\cite{Booker2020,Dutta2025}.

However, the role of odd-even effects in connecting Liouvillian spectra to relaxation dynamics has received comparatively less attention. Focusing on the thermodynamic limit or on smooth finite-size scaling does not by itself resolve the distinction between odd and even chain lengths, while an analysis of spectral and initial-state contributions at a given size does not directly address how adding one site changes the coupling to the boundaries. Previous work considered a boundary-dephased XY chain and showed that the relaxation timescales obey different scaling laws for odd and even chain lengths in the weak-dissipation limit, with the difference washed out by strong dissipation~\cite{Zhang2020}. Odd-even differences have also been reported in the steady-state density profiles of boundary-driven double spin chains~\cite{Lingenfelter2024}. Exact persistent Liouvillian modes have been constructed in boundary-dissipative spin chains~\cite{Matsui2024}. Nevertheless, a criterion, in terms of the parity of chain length, for the existence of an entire exactly non-decaying operator sector, under fixed bulk couplings and a fixed boundary dissipation rule, has not been established. It therefore remains unclear whether changing the chain length from odd to even can control whether persistent oscillations exist, and what physical mechanism underlies this odd-even distinction.

In this work, we investigate odd-even effects on Liouvillian spectra and relaxation dynamics in boundary-driven XY and cluster chains. We show that changing the chain length alters the connectivity between the bulk degrees of freedom and the dissipative boundaries, thereby determining whether an entire operator sector remains decoupled from dissipation. The same endpoint spin flip channels lead to opposite odd-even effects in the two models: non-decaying oscillatory modes occur in even XY chains and odd cluster chains of sufficient length. This distinction is reflected in the presence or absence of nonzero purely imaginary Liouvillian eigenvalues and can be observed through bond correlations and magnetization following simple product-state preparations. We identify the underlying mechanism through the connectivity of Majorana paths to the dissipative contacts and extend it to cluster interactions of even range under a fixed boundary dissipation rule. The existence and frequencies of the persistent modes are independent of the magnitudes of the positive boundary dissipation rates, including unequal rates. Our results establish the parity of chain length as a discrete control of dissipation-free dynamics and connect odd-even effects in the spectrum directly to measurable relaxation behavior.

The paper is organized as follows. Section~\ref{sec:xy} presents the XY chain, its Liouvillian spectrum, and spin correlation dynamics. Section~\ref{sec:cluster} establishes the reversed odd-even effect in cluster chains and its magnetization signature. Section~\ref{sec:conclusions} summarizes the results.

\section{Boundary-driven XY chain}\label{sec:xy}

\subsection{Model}

We begin with a zero-field XY chain of $N\ge2$ spins, with Pauli operators $\sigma_j^\zeta$ ($\zeta=x,y,z$) and real exchange couplings,
\begin{equation}
H_{\rm XY}=-\sum_{j=1}^{N-1}(J_x\sigma_j^x\sigma_{j+1}^x+J_y\sigma_j^y\sigma_{j+1}^y).
\label{eq:Hxy}\end{equation}
Here we take $J_xJ_y\ne0$ and $\hbar=1$. The density matrix $\rho(t)$ obeys the Lindblad master equation with Liouvillian $\LL$~\cite{Gorini1976,Lindblad1976}
\begin{align}
&\dot\rho=\LL[\rho]=-\ii[H,\rho]+\sum_\alpha\D[L_\alpha]\rho,\label{eq:lindblad}\\
&\D[L]\rho=L\rho L^\dagger-\tfrac12\{L^\dagger L,\rho\}.\nonumber
\end{align}

The two boundary jump operators are
\begin{equation}
L_L=\sqrt{\gamma_L}\sigma_1^x,\qquad L_R=\sqrt{\gamma_R}\sigma_N^x,\quad \gamma_L,\gamma_R>0.\label{eq:jumps}
\end{equation}
Both jumps are Hermitian, so $\LL[I]=0$, and $I/2^N$ is a steady state.

Define the Hermitian Majorana operators by
\begin{equation}\label{eq:physicalJW}
a_j=\left(\prod_{k<j}\sigma_k^z\right)\sigma_j^x,
b_j=\left(\prod_{k<j}\sigma_k^z\right)\sigma_j^y.
\end{equation}
In the ordering $w=(a_1,b_1,\ldots,a_N,b_N)^T$, these operators obey $\{w_m,w_n\}=2\delta_{mn}I$. We also define the spin parity $P_s=\prod_j\sigma_j^z$ and the operator parity map $\PP(O)=P_sOP_s$, with eigenvalue $P=\pm1$. Here operator parity is distinct from the odd-even character of the chain length or site indices. The Hamiltonian is even and the jumps odd under $P_s$, so the Liouvillian preserves operator parity~\cite{Buca2012}.

\begin{figure}[t]

\centering

\includegraphics[width=\columnwidth]{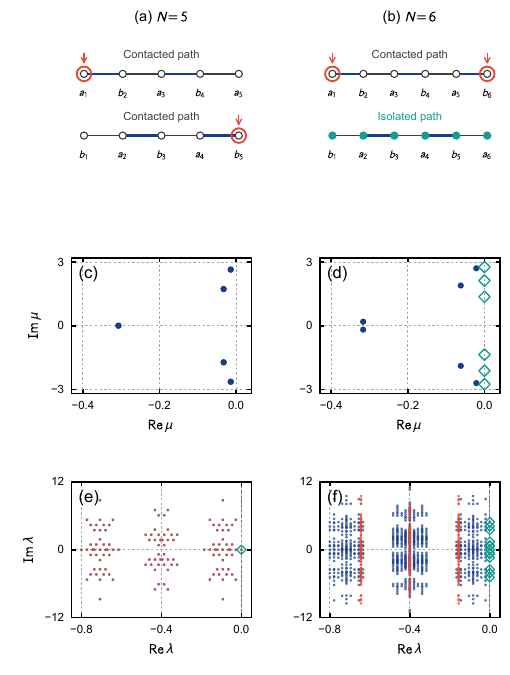}

\caption{XY chain at $J_x=1$, $J_y=0.5$, and $\gamma_L=\gamma_R=0.2$, with $N=5,6$ in the left and right columns. (a,b) Majorana paths: blue and gray links denote $J_y$ and $J_x$, rings mark Majorana coordinates acted on by boundary dissipation, and green nodes are isolated. (c,d) Rapidity spectra. (e,f) Liouvillian spectra: blue and red denote the even and odd operator sectors. Green diamonds mark eigenvalues with zero real part.}\label{fig:xystructure}

\end{figure}

Here a Majorana path denotes a sequence of Majorana operators coupled by the quadratic Hamiltonian; an isolated path is one on which boundary dissipation does not act in the even operator subspace. The Hamiltonian couples $b_j$ to $a_{j+1}$ and $a_j$ to $b_{j+1}$, forming two independent Majorana paths, $\chi_1=(a_1,b_2,a_3,\ldots)$ and $\chi_2=(b_1,a_2,b_3,\ldots)$. The left jump is proportional to $a_1$, while $\sigma_N^x=\ii P_s b_N$ at the right boundary generates the same dissipative action as $b_N$ on even-parity operators. For even $N$, both boundary dissipators act on $\chi_1$ in the even operator subspace, leaving $\chi_2$ isolated from dissipation, as shown in Fig.~\ref{fig:xystructure}(b). For odd $N$, boundary dissipation acts on both paths, leaving neither path isolated, as shown in Fig.~\ref{fig:xystructure}(a). The explicit Majorana couplings are given in Appendix~\ref{app:xy-paths}. This odd-even dependence of the action of boundary dissipation accounts for the distinct Liouvillian spectra and long-time dynamics.

\subsection{Liouvillian spectrum}

We write the quadratic Hamiltonian as $H=(\ii/4)w^TAw$, where $w=(w_1,\ldots,w_{2N})^T$ is the vector of Majorana operators and $A$ is a real antisymmetric matrix. The effect of boundary dissipation is encoded in the positive semidefinite damping matrix $D$, giving the rapidity spectrum matrix $M=A-2D$. For the endpoint channels considered here,

\begin{equation}
M=A-2D,\qquad D=\gamma_LE_1^{(2N)}+\gamma_RE_{2N}^{(2N)},
\label{eq:M}
\end{equation}
where $E_j^{(2N)}=e_je_j^T$ and $e_j$ is the $j$th standard coordinate vector. We denote the eigenvalues of $M$ by $\mu$ (rapidities) and those of the Liouvillian by $\lambda$. The matrix $M$ determines the evolution of even Majorana bilinears, as shown below; its eigenvalues are not themselves the full Liouvillian spectrum.

For an eigenmode satisfying $Mv=\mu v$, the antisymmetry of $A$ yields

\begin{equation}
\operatorname{Re}\mu=-2\frac{v^\dagger Dv}{v^\dagger v}\leq 0.
\label{eq:contact-decay}
\end{equation}

Hence, an undamped mode must satisfy $Dv=0$ and $Av=\mu v$. This condition leads to a simple criterion for undamped modes: for independent finite open Majorana paths with nonzero couplings between adjacent Majorana coordinates and independent dissipation with positive rates acting only at selected endpoints, a path supports undamped modes if and only if neither endpoint is coupled to a dissipative channel. Indeed, on any path subject to boundary dissipation, $Dv=0$ forces the amplitude at the damped endpoint to vanish, and the eigenvalue equation then recursively forces all amplitudes along that path to vanish. By contrast, an isolated path retains its purely imaginary spectrum because its rapidity spectrum matrix is real and antisymmetric. Consequently, the number $d$ of undamped rapidity modes equals the total number of Majorana coordinates on the isolated paths. The criterion for undamped modes is derived in Appendix~\ref{app:contact}; its connection to non-decaying spin operators is discussed in Appendix~\ref{app:boundary-action}.

For even $N$, the isolated XY path contains $N$ Majorana coordinates, whereas neither path is isolated for odd $N$. For $N\geq 2$, $J_xJ_y\neq 0$, and positive boundary dissipation rates in Eq.~\eqref{eq:jumps}, the criterion for undamped modes yields

\begin{equation}
d_{\rm XY}=\begin{cases}0,&N\text{ odd},\\N,&N\text{ even}.\end{cases}
\label{eq:xy-d}
\end{equation}
Liouvillian eigenvalues on the imaginary axis correspond to non-decaying modes. Eigenoperators satisfying $\LL[R]=\lambda R$ with $\operatorname{Re}\lambda=0$ span the \emph{non-decaying operator subspace}: zero eigenvalues describe stationary dynamics, while the imaginary parts of nonzero purely imaginary eigenvalues determine persistent oscillation frequencies. We reconstruct the physical Liouvillian spectrum using the method with projection onto operator parity sectors of Ref.~\cite{Zheng2023}, with the operator parity constraint in Eq.~\eqref{eq:spectrum} and the XY blocks in Eq.~\eqref{eq:XYblocks}.

Figures~\ref{fig:xystructure}(c,d) show the rapidity spectra, and Figs.~\ref{fig:xystructure}(e,f) show the full Liouvillian spectra, for $N=5,6$ at $J_x=1$, $J_y=0.5$, and $\gamma_L=\gamma_R=0.2$. For odd $N=5$, both paths are subject to boundary dissipation and every rapidity has a strictly negative real part in Fig.~\ref{fig:xystructure}(c). For even $N=6$, one path is isolated and six rapidities lie exactly on the imaginary axis in Fig.~\ref{fig:xystructure}(d). Their number agrees with $d_{\rm XY}=N$ in Eq.~\eqref{eq:xy-d}. The remaining rapidities have negative real parts and describe damped modes.

The full Liouvillian spectra show the corresponding distinction between stationary and oscillatory dynamics. For odd $N=5$, only stationary modes remain non-decaying in Fig.~\ref{fig:xystructure}(e). There are no nonzero purely imaginary eigenvalues. For even $N=6$, nonzero purely imaginary eigenvalues appear alongside the stationary modes at the origin in Fig.~\ref{fig:xystructure}(f). They supply non-decaying oscillatory modes, while eigenvalues with negative real parts describe decaying contributions. A nonzero purely imaginary eigenvalue contributes a persistent oscillation only when the total coefficient at that frequency, determined by the initial state and measured observable, is nonzero. The spin correlation dynamics below provides this connection.

\begin{figure}[t]

\centering

\includegraphics[width=0.8\columnwidth]{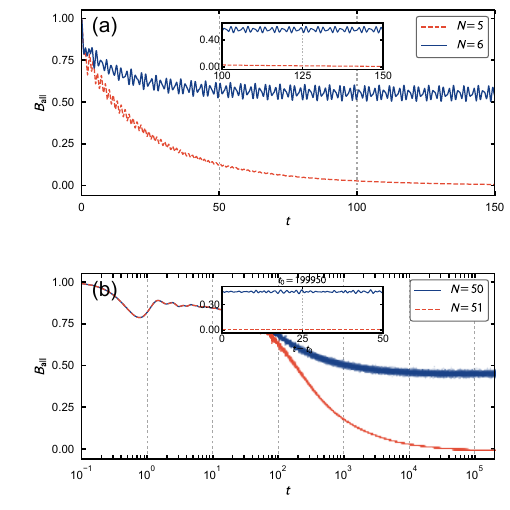}

\caption{XY spin correlation dynamics from $\ket{+x}^{\otimes N}$ at $J_x=1$, $J_y=0.5$, and $\gamma_L=\gamma_R=0.2$. (a) $N=5,6$. Inset: $100\le t\le150$. (b) $N=50,51$. Inset: $199950\le t\le200000$. Blue solid/red dashed curves denote even/odd lengths.}\label{fig:xy}

\end{figure}

\subsection{Spin correlation dynamics}

The magnetization and bond observables studied here are even under $\PP$. Their expectations depend only on $\rho_+=(\rho+P_s\rho P_s)/2$, since $\Tr[O\rho_-]=0$ for even $O$ and $\rho_-=(\rho-P_s\rho P_s)/2$. We therefore obtain their dynamics from the even bilinears. With $\langle O\rangle_t=\Tr[\rho(t)O]$, define the covariance matrix and its evolution by~\cite{Prosen2008,Barthel2022,Landi2022}
\begin{equation}
C_{mn}=\frac{\ii}{2}\Tr(\rho[w_m,w_n]).\label{eq:C}
\end{equation}
\begin{equation}
\dot C=MC+CM^T,\qquad C(t)=e^{Mt}C(0)e^{M^Tt},\label{eq:covdyn}
\end{equation}
A local $\sigma_j^z$ connects the two XY paths, whereas a bond correlation can probe one path directly. For even chains, odd-indexed $xx$ bonds access bilinears within the isolated path. Using the covariance evolution in Eq.~\eqref{eq:covdyn}, we prepare $\ket{\psi_0}=\ket{+x}^{\otimes N}$, with $\sigma^x\ket{+x}=\ket{+x}$, and measure
\begin{equation}
B_{\rm all}(t)=\frac1{N-1}\sum_{j=1}^{N-1}\langle\sigma_j^x\sigma_{j+1}^x\rangle_t
=-\frac1{N-1}\sum_{j=1}^{N-1}C_{2j,2j+1}(t).
\label{eq:Ball}
\end{equation}
Initially the independent nonzero covariances are $C_{2j,2j+1}=-1$, so $B_{\rm all}(0)=1$.

For even $N\geq4$, odd-indexed bonds lie within the isolated path $\chi_2$ and even-indexed bonds lie within the path subject to boundary dissipation $\chi_1$. Define $B_\eta(t)=n_\eta^{-1}\sum_{j\in\eta}\langle\sigma_j^x\sigma_{j+1}^x\rangle_t$, where $\eta=\mathrm{odd},\mathrm{even}$ specifies whether the bond's left site has an odd or even index, $n_{\rm odd}=N/2$, and $n_{\rm even}=(N-2)/2$. Then

\begin{equation}
B_{\rm all}(t)=\frac{N}{2(N-1)}B_{\rm odd}(t)
+\frac{N-2}{2(N-1)}B_{\rm even}(t), (N\text{ even}).
\label{eq:Bdecomp}
\end{equation}
For $N=50$ the coefficients are $25/49$ and $24/49$. For even $N\geq4$, the first term is independent of the boundary dissipation rates, whereas the second decays. At $N=2$ there is only one bond. Its average is $B_{\rm odd}$ and no $B_{\rm even}$ average is defined. For odd $N$ both bond components decay.

Figure~\ref{fig:xy}(a) shows the dynamical consequence of the $N=5,6$ spectra at the same parameters: odd-chain bond correlations decay, while the even chain retains an oscillatory contribution. The $N=50,51$ comparison in (b) exhibits the same non-decaying dynamics for even-length chains. The logarithmic-time comparison reaches $t=2\times10^5$. In its final linear-axis inset, the transient envelope is below $10^{-4}$ for both sizes. The even-chain oscillations persist while the odd-chain bond correlation has relaxed.

For the XY chain with $N=4$, the non-decaying contribution has the explicit form

\begin{equation}
B_{\rm odd}(t)=\frac{4J_x^2+J_y^2\cos(2\sqrt{4J_x^2+J_y^2}\,t)}{4J_x^2+J_y^2}.
\label{eq:minxy}\end{equation}
It exhibits both a constant contribution and a contribution at a nonzero frequency, each independent of the boundary dissipation rates. At $N=2$ the isolated component instead supports only stationary bond correlations. For $N=4$, the correlation averaged over the chain is $B_{\rm all}=(2/3)B_{\rm odd}+(1/3)B_{\rm even}$, with the second term decaying. The same covariance construction gives the $N=6$ curves in Fig.~\ref{fig:xy}(a). Its effective two-level derivation is in Appendix~\ref{app:minimal}. Results for $J_y=1.5$ are given in Appendix~\ref{app:regime}.

The same boundary jump operators can reverse which chain lengths support non-decaying dynamics when the bulk Hamiltonian couples the Majorana operators differently. We demonstrate this reversal in the cluster chain.

\section{Boundary-driven cluster chains}\label{sec:cluster}

\subsection{Model}

Cluster interactions provide a second realization of the mechanism of isolation from boundary dissipation, with an odd-even effect opposite to that of the XY chain. Cluster states are resources for measurement-based quantum computation~\cite{Raussendorf2001}, and transverse-field cluster Hamiltonians can be reduced to two Ising sectors~\cite{Doherty2009}. Long-lived boundary qubits have also been studied in perturbed, dimerized cluster chains~\cite{Kemp2020}. We consider a cluster family with even interaction ranges and use the cluster chain with $r=2$ to relate the Majorana paths to the Liouvillian spectrum and magnetization dynamics. The open-chain Hamiltonian is

\begin{equation}
H_r=-J\sum_{j=1}^{N-r}\sigma_j^x
 \left(\prod_{p=j+1}^{j+r-1}\sigma_p^z\right)\sigma_{j+r}^x
 -h\sum_{j=1}^{N}\sigma_j^z,
\label{eq:Hr-family}
\end{equation}
where $r=2s$ is a fixed even interaction range, $s$ is a positive integer, $J$ is the cluster coupling, $h$ is the transverse field, $N\ge r$, and $Jh\ne0$. The string has $r-1$ intermediate spins. The boundary construction follows the $r$ paths generated by the bulk coupling, each linking sites separated by $r$. For even $r$, each path contains only odd-indexed or only even-indexed sites. Applying dissipation to alternating endpoint Majorana coordinates therefore acts on one half of the paths at each boundary. We use the same offsets from the left and right ends for every allowed $N$, without changing the form of the boundary jump operators when the length changes. In the Lindblad equation~\eqref{eq:lindblad}, the independent boundary channels are, for $k=0,\ldots,s-1$,

\begin{align}
L_{L,k}&=\sqrt{\gamma_{L,k}}
 \left(\prod_{p=1}^{2k}\sigma_p^z\right)\sigma_{2k+1}^x,\nonumber\\
L_{R,k}&=\sqrt{\gamma_{R,k}}\sigma_{N-2k}^x
 \left(\prod_{p=N-2k+1}^{N}\sigma_p^z\right).
\label{eq:family-jumps}
\end{align}

All rates are positive and empty products are the identity. The channels enter independently in the dissipator sum. For fixed $r$, the jumps are supported within $r-1$ sites of each end. At $r=2$, they reduce to the same two endpoint channels in Eq.~\eqref{eq:jumps}. For each fixed $r$, there are $r/2$ channels per end: their spin strings produce dissipation on individual endpoint Majorana coordinates in the even operator subspace. As $N$ changes, the Majorana coordinates acted on by the right boundary dissipators follow the same rule relative to the chain ends, while their assignment to the bulk paths changes. This is what produces the odd-even effect.

The cluster chain with $r=2$ has the Hamiltonian

\begin{equation}
H_{\rm cl}=-J\sum_{j=1}^{N-2}\sigma_j^x\sigma_{j+1}^z\sigma_{j+2}^x-h\sum_{j=1}^N\sigma_j^z,
\label{eq:Hcl}\end{equation}
with the channels in Eq.~\eqref{eq:jumps}. Under the Majorana transformation~\eqref{eq:physicalJW}, the Hamiltonian for $r=2$ becomes

\begin{equation}
H_{\rm cl}=\ii h\sum_{j=1}^N a_jb_j+\ii J\sum_{j=1}^{N-2}b_ja_{j+2}.\label{eq:physicalcl}
\end{equation}
More generally, the interaction in $H_r$ couples $b_j$ to $a_{j+r}$. The Majorana graph separates into $r$ paths, each containing sites with the same index modulo $r$. Because $r$ is even, each path contains physical sites whose indices are all odd or all even. This structure allows the action of the boundary channels to be determined before diagonalizing the generator.

\begin{figure}[t]

\centering

\includegraphics[width=\columnwidth]{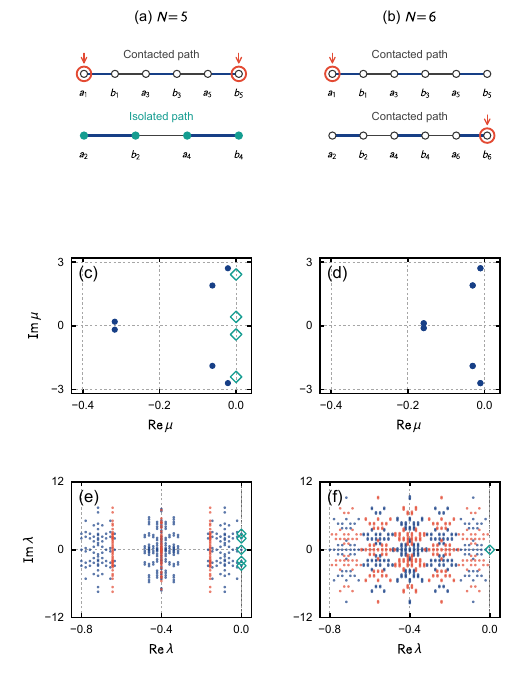}

\caption{Cluster chain at $r=2$, $J=1$, $h=0.5$, and $\gamma_L=\gamma_R=0.2$, with $N=5,6$ in the left and right columns. (a,b) Majorana paths: blue and gray links denote $h$ and $J$, red rings mark Majorana coordinates acted on by boundary dissipation, and green nodes are isolated. (c,d) Rapidity spectra. (e,f) Liouvillian spectra.}\label{fig:clstructure}

\end{figure}

\subsection{Liouvillian spectrum}

In the even operator sector, the boundary dissipators act on $a_{2k+1}$ and $b_{N-2k}$, respectively, since $L_{R,k}=\ii\sqrt{\gamma_{R,k}}P_sb_{N-2k}$. The corresponding damping matrix is $D_r$ in Eq.~\eqref{eq:family-drift}.

The left boundary dissipators act on the endpoints of all paths on odd sites. The right boundary dissipators act on the same paths for odd $N$, leaving the even-site paths isolated, but the complementary paths for even $N$. The criterion for undamped modes therefore gives

\begin{equation}
d_{\rm cl}=\begin{cases}N-1,&N\text{ odd},\\0,&N\text{ even}.\end{cases}
\label{eq:family-d}
\end{equation}

For odd $N$, the isolated even sites contain $2\lfloor N/2\rfloor=N-1$ Majorana operators. For even $N$, every path is subject to boundary dissipation and all rapidity modes decay. This reverses the odd-even effect in the XY chain despite the identical endpoint jumps in the XY chain and the cluster chain with $r=2$. Whether odd or even chains support undamped modes therefore depends on how the boundary jump operators act on the Majorana paths generated by the bulk Hamiltonian. Importantly, the isolated-path block of the rapidity matrix is independent of the magnitudes of the positive boundary dissipation rates. Varying these rates modifies the spectra of the boundary-coupled paths but leaves the undamped rapidities and the corresponding purely imaginary Liouvillian eigenvalues unchanged.

We now specialize to $r=2$ and compare the rapidity and full Liouvillian spectra in Fig.~\ref{fig:clstructure}. The full spin spectrum is reconstructed using Eq.~\eqref{eq:spectrum} and the cluster blocks in Eq.~\eqref{eq:CLblocks}. Their derivation is given in Appendix~\ref{app:cluster}.

Figures~\ref{fig:clstructure}(c,d) show the rapidity spectra, and Figs.~\ref{fig:clstructure}(e,f) show the full Liouvillian spectra, for $N=5,6$ at $J=1$, $h=0.5$, and $\gamma_L=\gamma_R=0.2$. In the rapidity spectra, the odd chain has four purely imaginary eigenvalues in Fig.~\ref{fig:clstructure}(c), exactly as predicted by $d_{\rm cl}=N-1=4$ in Eq.~\eqref{eq:family-d}. The even chain has no imaginary-axis rapidities: every mode has a strictly negative real part in Fig.~\ref{fig:clstructure}(d). This reverses the odd-even dependence of the XY rapidities under the same endpoint channels.

The full Liouvillian spectra display the corresponding reversal. At odd $N=5$, nonzero purely imaginary eigenvalues coexist with stationary modes at the origin in Fig.~\ref{fig:clstructure}(e), supporting persistent oscillations. At even $N=6$, only stationary modes remain non-decaying in Fig.~\ref{fig:clstructure}(f), and all nonstationary modes decay. The magnetization dynamics below probes the persistent frequencies of the odd chain. The corresponding spectra and dynamics for $r=4,6$ are presented in Appendix~\ref{app:family}. The robustness of this parity-dependent spectral and dynamical behavior against unequal positive boundary dissipation rates is demonstrated in Appendix~\ref{app:unequal}.

\begin{figure}[t]

\centering

\includegraphics[width=0.8\columnwidth]{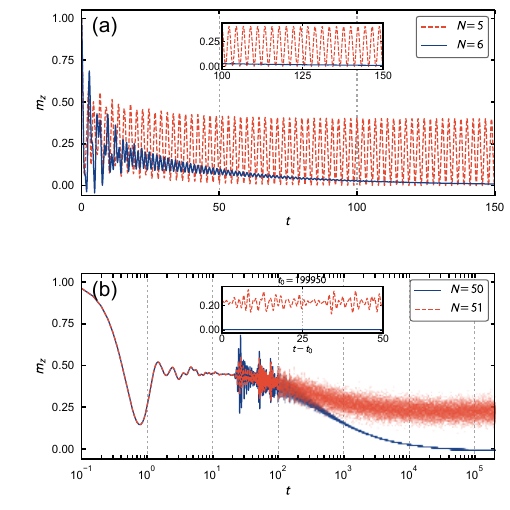}

\caption{Cluster magnetization from $\ket{\uparrow_z}^{\otimes N}$ at $J=1$, $h=0.5$, and $\gamma_L=\gamma_R=0.2$. (a) $N=5,6$. Inset: $100\le t\le150$. (b) $N=50,51$. Inset: $199950\le t\le200000$. Blue solid/red dashed curves denote even/odd lengths.}\label{fig:cluster}

\end{figure}

\subsection{Magnetization dynamics}

We first express the magnetization dynamics for the even-range family using Eq.~\eqref{eq:covdyn} with $M_r=A_r-2D_r$, and then evaluate the $r=2$ chain shown in Fig.~\ref{fig:cluster}.

An on-site magnetization probes the cluster paths directly because $a_j$ and $b_j$ lie on the same path. We prepare $\ket{\psi_0}=\ket{\uparrow_z}^{\otimes N}$, with $\sigma^z\ket{\uparrow_z}=\ket{\uparrow_z}$, and measure

\begin{equation}
m_z(t)=\frac1N\sum_{j=1}^N\langle\sigma_j^z\rangle_t
=-\frac1N\sum_{j=1}^NC_{2j-1,2j}(t).
\label{eq:mz}
\end{equation}

The initial covariance has $C_{2j-1,2j}=-1$ and no other independent nonzero entries. Define $m_\eta(t)=N_\eta^{-1}\sum_{j\in\eta}\langle\sigma_j^z\rangle_t$, where $\eta=\mathrm{odd},\mathrm{even}$ labels odd- and even-indexed sites, $N_{\rm odd}=\lceil N/2\rceil$, and $N_{\rm even}=\lfloor N/2\rfloor$. For odd $N$, these averages satisfy

\begin{equation}
m_z(t)=\frac{N+1}{2N}m_{\rm odd}(t)+\frac{N-1}{2N}m_{\rm even}(t),\label{eq:mdecomp}
\end{equation}
where $m_{\rm even}$ evolves independently of dissipation for every fixed even $r$ with the boundary jump operators in Eq.~\eqref{eq:family-jumps}. For even $N$, all paths are subject to boundary dissipation and both sublattice magnetizations decay. The following explicit dynamical examples use $r=2$. At $N=51$, the two weights are $26/51$ and $25/51$.

Figure~\ref{fig:cluster}(a) uses exactly the same sizes and parameters as the spectral comparison. The odd $N=5$ chain retains oscillations, whereas the even $N=6$ signal decays, directly linking non-decaying Liouvillian modes to a measurable spin response. Figure~\ref{fig:cluster}(b) extends the comparison to $N=50,51$. The small-chain window already resolves the oscillatory contrast. For the larger chains, the logarithmic axis follows the transient to $t=2\times10^5$, and the final linear-axis inset resolves the remaining oscillations. At that window the estimated absolute transient envelope is below $10^{-4}$ for both curves, so the odd-chain signal is governed by its non-decaying Liouvillian modes rather than a slowly decaying contribution (see Appendix~\ref{app:covariance}). The non-decaying contribution is carried by the even sublattice and is independent of the boundary dissipation rates. The corresponding $h=1.5$ dynamics is shown in Appendix~\ref{app:cluster-regime} with the same layout.

At $N=5$, the even-sublattice magnetization is

\begin{equation}
m_{\rm even}(t)=\frac{4h^2+J^2\cos(2\sqrt{4h^2+J^2}\,t)}{4h^2+J^2}.
\label{eq:mincl}\end{equation}

This expression contains both a stationary part and an oscillatory part. It describes the even-site sublattice decoupled from dissipation, whereas Fig.~\ref{fig:cluster}(a) shows the average over the whole chain: at $N=5$, Eq.~\eqref{eq:mdecomp} gives $m_z=(3/5)m_{\rm odd}+(2/5)m_{\rm even}$, with $m_{\rm odd}\to0$ at long times. At $N=3$, however, the spin observable $\sigma_2^z$ is conserved, so the signal retains memory without oscillating. The derivation is given in Appendix~\ref{app:cluster-dynamics}. For larger even ranges, the isolated paths obey the same dynamics with alternating couplings. Together with the XY bond response, this magnetization signal shows how different bulk paths determine both which chain lengths support non-decaying dynamics and which spin observables detect it.

\section{Conclusions}\label{sec:conclusions}
We have investigated odd-even effects on Liouvillian spectra and relaxation dynamics in boundary-driven XY and cluster chains. With the bulk couplings and the boundary-dissipation rule fixed, the parity of the chain length determines which Majorana paths couple to the boundary jump operators. An isolated path supports undamped modes, and products of an even number of Majorana operators along such a path span a non-decaying subspace of even-parity operators. The same endpoint spin-flip channels therefore lead to opposite parity-dependent behavior in the two models. For sufficiently long chains, even-length XY chains and odd-length cluster chains with $r=2$ support nonzero purely imaginary Liouvillian eigenvalues, giving rise to persistent oscillations in bond correlations and magnetization, respectively.

These results connect odd-even effects to the existence of non-decaying oscillatory modes and extend the discussion beyond parity effects on relaxation timescales. We establish a criterion for undamped modes that moves beyond previous case-specific constructions of persistent modes and decoherence-free dynamics, showing precisely when a fixed boundary-dissipation rule isolates bulk degrees of freedom. The same mechanism applies to even-range cluster interactions with the prescribed boundary channels. Varying the positive boundary dissipation rates, including making them unequal, modifies the decay rates of boundary-damped modes but leaves the isolated paths unchanged. Consequently, the existence and frequencies of the persistent modes are independent of the magnitudes of the boundary dissipation rates, provided all prescribed rates remain positive. Thus, the parity of the chain length provides discrete control over non-decaying dynamics, with signatures accessible through state preparation in product states and spin measurements.
 
{\bf Acknowledgments.} S.C. was supported by
the National Key Research and Development Program of
China (Grant No. 2023YFA1406704) and the NSFC under
Grants No. 12474287 and No. T2121001.

{\bf Data avail ability.}
The data that support the findings of this article are openly
available{\cite{zhengdata}}.

\appendix

\section{XY chain}\label{app:xy}

\subsection{Majorana representation and boundary dissipation}\label{app:xy-paths}\label{app:contact}\label{app:boundary-action}

Writing $\chi_{\nu,j}$ for the $j$th Majorana on path $\chi_\nu$ defined in Sec.~\ref{sec:xy}, the physical decomposition is

\begin{equation}
\begin{aligned}
H_{\rm XY}&=\sum_{\nu=1}^2\ii\sum_{j=1}^{N-1}\kappa_{\nu,j}\chi_{\nu,j}\chi_{\nu,j+1},\\
(\kappa_{1,j},\kappa_{2,j})&=\begin{cases}(-J_y,J_x),&j\text{ odd},\\(J_x,-J_y),&j\text{ even}.
\end{cases}
\end{aligned}\label{eq:physicalxy}
\end{equation}

For odd $N$, each path contains an odd number of physical Majoranas, so its operators cannot be paired into an independent set of complex fermions. The two paths must therefore be retained when reconstructing the physical spin Liouvillian spectrum.

For $M=A-2D$, with real antisymmetric $A$ and $D=D^\dagger\ge0$, multiplying $Mv=\mu v$ by $v^\dagger$ and taking the real part gives Eq.~\eqref{eq:contact-decay}. Positive semidefiniteness implies $v^\dagger Dv=0$ precisely when $Dv=0$. Hence, for any eigenpair with $v\ne0$,

\begin{equation}
\operatorname{Re}\mu=0
\quad\Longleftrightarrow\quad
Dv=0\ \text{and}\ Av=\mu v.
\label{eq:undamped-criterion}
\end{equation}

Conversely, an eigenvector of $A$ satisfying $Dv=0$ is unaffected by damping and is also an undamped eigenvector of $M$.

For independent finite open paths with nonzero adjacent couplings and diagonal damping with positive rates at selected endpoints, $Dv=0$ sets the amplitude at each endpoint subject to dissipation to zero. The endpoint eigenvalue equation then sets the neighboring amplitude to zero, and recursion eliminates the entire path subject to boundary dissipation. An isolated path has an antisymmetric block of the rapidity spectrum matrix and purely imaginary eigenvalues. This establishes the criterion for undamped modes used in the main text.

For the specified boundary jump operators, products of an even number of isolated Majorana operators commute with all jumps, and the Hamiltonian preserves the subspace they span. Within each even-degree Majorana subspace, the eigenvalues of the adjoint Liouvillian are sums of distinct rapidities. Products of isolated eigenmodes therefore evolve without decay, whereas sums containing a damped eigenmode have negative real part. Zero and nonzero imaginary sums give stationary and oscillatory contributions, respectively. This construction concerns even operators; the full spin Liouvillian spectrum also requires the operator parity projection derived below.

The absence of decay in the even operator subspace follows without replacing the full right jump by a linear fermion. For an operator satisfying $P_sOP_s=PO$, with $P=\pm1$,

\begin{equation}
\begin{aligned}
\sigma_N^xO\sigma_N^x&=P\,b_NOb_N,\\
\D^\dagger[L_R]O&=\gamma_R(P\,b_NOb_N-O).
\end{aligned}\label{eq:exactright}
\end{equation}

Here $\D^\dagger$ is the adjoint dissipator acting on observables. For even XY chains, even products of Majoranas on the isolated path commute with both boundary jumps and evolve within that path. Equation~\eqref{eq:exactright} also explains why a linear-Majorana dissipator is adequate for the even bilinears, but not for arbitrary parity-odd spin operators.

\subsection{Liouvillian spectrum}\label{app:xy-spectrum}

For the XY chain, fermionization in the doubled space in a fixed sector gives~\cite{Zheng2023}

\begin{equation}
\begin{aligned}
\LL_P&=2\ii c^\dagger T^P c+\gamma_L(P-1),\\
\PP&=(-1)^{\hat N_f},\qquad
\hat N_f=\sum_{\ell=1}^{2N}c_\ell^\dagger c_\ell.
\end{aligned}\label{eq:LP}
\end{equation}
Here $c=(c_1,\ldots,c_{2N})^T$ contains $2N$ auxiliary complex fermions with $\{c_\ell,c_m^\dagger\}=\delta_{\ell m}$ and $\{c_\ell,c_m\}=0$. The $2N\times2N$ auxiliary rapidity spectrum matrix $T^P$ is complex symmetric. In this representation, $\PP$ is the matrix of the operator parity superoperator defined in Sec.~\ref{sec:xy}. The XY blocks are given in Eq.~\eqref{eq:XYblocks}. Let $E_{\nu,\ell,P}$ be their eigenvalues, with $\nu$ labeling a block and $\ell$ a mode within that block. These auxiliary rapidities are distinguished from the rapidities $\mu$ of $M$: in the even operator sector, $M$ is similar to $2\ii T^+$, so $\mu_k=2\ii E_{k,+}$, as derived in Eq.~\eqref{eq:similarity}. The physical spin spectrum is

\begin{equation}
\begin{gathered}
\lambda_P=2\ii\sum_{\nu,\ell}n_{\nu,\ell}E_{\nu,\ell,P}+\gamma_L(P-1),\\
n_{\nu,\ell}\in\{0,1\},\qquad (-1)^{\sum_{\nu,\ell}n_{\nu,\ell}}=P.
\end{gathered}\label{eq:spectrum}
\end{equation}

The numbers $n_{\nu,\ell}$ are mode occupations. The constraint is imposed on their total across both blocks. The vectorization below derives both the parity constraint and the shift determined by the sector.

With vectorization by rows $\ket{\rho}\!\rangle=\sum_{mn}\rho_{mn}\ket m\otimes\ket n$, the map $X\rho Y$ becomes $X\otimes Y^T$ for operators $X,Y$. Denote replica spins by $\sigma_{j,L}^\zeta=\sigma_j^\zeta\otimes I$ and $\sigma_{j,R}^\zeta=I\otimes\sigma_j^\zeta$, with $\zeta=x,y,z$. Then

\begin{equation}
\begin{aligned}
\LL={}&\ii(I\otimes H^T-H\otimes I)
+\sum_\alpha\Big[L_\alpha\otimes L_\alpha^*
 \\&-\tfrac12 L_\alpha^\dagger L_\alpha\otimes I
-\tfrac12 I\otimes L_\alpha^TL_\alpha^*\Big].
\end{aligned}\label{eq:vec}
\end{equation}

Vectorization also underlies the non-Hermitian Kitaev ladder and Ashkin--Teller representations of other dissipative spin chains~\cite{ShibataKitaev2019,ShibataIsing2019}. Those mappings involve different bulk jumps and sector structures. The boundary identities below are specific to the present models.

For the XY Hamiltonian, $H^T=H$. The dissipative part is $\gamma_L\sigma_{1,L}^x\sigma_{1,R}^x+\gamma_R\sigma_{N,L}^x\sigma_{N,R}^x-(\gamma_L+\gamma_R)I$.

Define $S_{L,<j}=\prod_{k<j}\sigma_{k,L}^z$, $S_{R,>j}=\prod_{k>j}\sigma_{k,R}^z$, and $P_L=\prod_k\sigma_{k,L}^z$. To distinguish auxiliary and physical Majoranas, use

\begin{equation}
\begin{aligned}
\mathfrak a_j&=(-1)^jS_{L,<j}\sigma_{j,L}^x,
\mathfrak b_j=(-1)^{j+1}S_{L,<j}\sigma_{j,L}^y,\\
\bar{\mathfrak a}_j&=(-1)^jP_LS_{R,>j}\sigma_{j,R}^x,
\bar{\mathfrak b}_j=(-1)^jP_LS_{R,>j}\sigma_{j,R}^y.
\end{aligned}\label{eq:auxJW}
\end{equation}
Distinct operators in this set of $4N$ Majoranas anticommute, and each squares to the identity. This explicit phase convention fixes the signs of the auxiliary hoppings and should be used together with the pairing

\begin{equation}
\begin{aligned}
\mathfrak a_j&=c_{2j-1}+c_{2j-1}^\dagger,
\bar{\mathfrak b}_j=(c_{2j-1}-c_{2j-1}^\dagger)/\ii,\\
\bar{\mathfrak a}_j&=c_{2j}+c_{2j}^\dagger,
\mathfrak b_j=(c_{2j}-c_{2j}^\dagger)/\ii.
\end{aligned}\label{eq:pairing}
\end{equation}
With $n_\ell=c_\ell^\dagger c_\ell$, the boundary identities are

\begin{equation}
\begin{aligned}
\sigma_{1,L}^x\sigma_{1,R}^x&=-2\PP(n_1-\tfrac12),\\
\sigma_{N,L}^x\sigma_{N,R}^x&=-2(n_{2N}-\tfrac12),\\
\PP&=(-1)^{\sum_\ell n_\ell}.
\end{aligned}\label{eq:boundaryaux}
\end{equation}
Consequently the constant is $P\gamma_L+\gamma_R-(\gamma_L+\gamma_R)=\gamma_L(P-1)$. There is no additional $P$ multiplying the right potential.

The number-conserving form of Eq.~\eqref{eq:LP} follows from the boundary identities above, as in Ref.~\cite{Zheng2023}. The reconstruction on physical operator space is

\begin{equation}
\LL=\sum_{P=\pm1}\Pi_P\LL_P\Pi_P,\qquad \Pi_P=\tfrac12(1+P\PP).\label{eq:projection}
\end{equation}

For the auxiliary Majorana vectors with entries $\alpha_\ell=c_\ell+c_\ell^\dagger$ and $\beta_\ell=(c_\ell-c_\ell^\dagger)/\ii$, the canonical anticommutation relations and $(T^P)^T=T^P$ give

\begin{equation}
c^\dagger T^P c=\tfrac12\Tr T^P+\tfrac{\ii}{2}\alpha^TT^P\beta.
\end{equation}
With $\xi=(\alpha^T,\beta^T)^T$, an explicit antisymmetric matrix form is

\begin{equation}
\begin{aligned}
\LL_P&=\tfrac14\xi^T\mathbb A_P\xi+\ii\Tr T^P+\gamma_L(P-1),\\
\mathbb A_P&=\begin{pmatrix}0&-2T^P\\2T^P&0\end{pmatrix},\qquad
\mathbb A_P^T=-\mathbb A_P.
\end{aligned}\label{eq:supermajorana}
\end{equation}

The matrix is generally complex, and $\LL_P$ is a generator rather than a Hermitian physical Hamiltonian. The constant must be retained: $\ii\Tr T^P+\gamma_L(P-1)=-(\gamma_L+\gamma_R)$ for the stated boundary potentials.

For a path with $n$ coordinates, define the alternating-hopping matrix

\begin{equation}
\begin{aligned}
S_n(u,v)&=\sum_{j=1}^{n-1}t_j(\ket{j}\bra{j+1}+\ket{j+1}\bra{j}),\\
t_j&=\begin{cases}u,&j\text{ odd},\\v,&j\text{ even}.
\end{cases}
\end{aligned}\label{eq:S}
\end{equation}

The corresponding coordinate projector is $E_j^{(n)}=\ket j\bra j$.

In the auxiliary space let $f_{1,j}=c_{2j-1}$ and $f_{2,j}=c_{2j}$ for odd $j$, interchanging the assignments for even $j$. The two blocks are

\begin{equation}
\begin{aligned}
T_1^P&=S_N(J_y,J_x)+\ii P\gamma_LE_1^{(N)}
                 +\ii(1-\delta_N)\gamma_RE_N^{(N)},\\
T_2^P&=S_N(J_x,J_y)+\ii\delta_N\gamma_RE_N^{(N)}.
\end{aligned}\label{eq:XYblocks}
\end{equation}

Here $\delta_N=[1-(-1)^N]/2$ equals one for odd $N$ and zero for even $N$. The left boundary is always $f_{1,1}$, whereas the right one is $f_{1,N}$ for even $N$ and $f_{2,N}$ for odd $N$. These statements determine the isolated path without inspecting numerical eigenvalues.

Substituting into Eq.~\eqref{eq:vec} gives

\begin{equation}
\begin{aligned}
\LL_{{\rm XY},P}={}&2\ii J_x\sum_{j=1}^{N-1}(c_{2j}^\dagger c_{2j+1}+\mathrm{H.c.})\\
&+2\ii J_y\sum_{j=1}^{N-1}(c_{2j-1}^\dagger c_{2j+2}+\mathrm{H.c.})\\
&-2P\gamma_Ln_1-2\gamma_Rn_{2N}+\gamma_L(P-1).
\end{aligned}\label{eq:auxxy}
\end{equation}

The nonzero off-diagonal entries of $T^P$ are $(T^P)_{2j,2j+1}=J_x$ and $(T^P)_{2j-1,2j+2}=J_y$, together with their transposes. The endpoint diagonal entries are $\ii P\gamma_L$ and $\ii\gamma_R$.

For diagonalizable $T^P$, write $T^PR_P=R_P\mathcal E_P$, where $\mathcal E_P$ is diagonal and $R_P$ is an invertible matrix of right eigenvectors. Define $\widehat f_{k,P}=\sum_\ell(R_P^{-1})_{k\ell}c_\ell$ and $\bar f_{k,P}=\sum_\ell c_\ell^\dagger(R_P)_{\ell k}$. Then $\{\widehat f_{k,P},\bar f_{\ell,P}\}=\delta_{k\ell}$ and $\sum_k\bar f_{k,P}\widehat f_{k,P}=\hat N_f$, although generally $\bar f_{k,P}\ne \widehat f_{k,P}^\dagger$. The projection $(-1)^{\hat N_f}=P$ remains necessary after diagonalization, yielding Eq.~\eqref{eq:spectrum}. The same eigenvalue formula remains valid by continuity when eigenvectors coalesce and the matrix is no longer diagonalizable.

\subsection{Spin correlation dynamics}\label{app:covariance}\label{app:minimal}

For $H=(\ii/4)w^TAw$, the Majorana anticommutation relations give $\ii[H,w_k]=\sum_lA_{kl}w_l$. The Hamiltonian contribution is consequently $\dot C|_H=AC+CA^T$. For $m\ne n$, conjugation of $w_mw_n$ by a boundary Majorana $w_q$ changes its sign exactly when one index equals $q$:

\begin{equation}
w_qw_mw_nw_q=(-1)^{\delta_{qm}+\delta_{qn}}w_mw_n.
\end{equation}

Using Eq.~\eqref{eq:exactright} in the even sector gives $\dot C_{mn}|_q=-2\gamma_q(\delta_{qm}+\delta_{qn})C_{mn}$. Together these results prove Eq.~\eqref{eq:covdyn}.

In the interleaved physical ordering, the independent nonzero Hamiltonian entries are

\begin{equation}
A_{2j,2j+1}=2J_x,\qquad A_{2j-1,2j+2}=-2J_y.\label{eq:Aentries}
\end{equation}
The opposite entries follow by antisymmetry, and all bonds are restricted to the chain. With $\mathsf J=\begin{pmatrix}0&1\\-1&0\end{pmatrix}$, the state $\ket{+x}^{\otimes N}$ has covariance
\begin{equation}
C_x(0)=0_{1\times1}\oplus\left[\bigoplus_{j=1}^{N-1}(-\mathsf J)\right]\oplus0_{1\times1}.\label{eq:initialC}
\end{equation}
It pairs $b_j$ with $a_{j+1}$, and the bond correlation is $\langle\sigma_j^x\sigma_{j+1}^x\rangle=-C_{2j,2j+1}$.

In the path basis, the covariance blocks obey $\dot C_{\nu\nu'}=M_\nu C_{\nu\nu'}+C_{\nu\nu'}M_{\nu'}^T$, where $\nu,\nu'$ label the paths.

For the XY chain, a phase transformation along each open path relates $M$ to the auxiliary rapidity spectrum matrix in the even operator sector:

\begin{equation}
M\simeq2\ii T^+,\qquad \mu_k=2\ii E_{k,+}.\label{eq:similarity}
\end{equation}
Here $k=1,\ldots,2N$ labels modes across both auxiliary blocks, so $E_{k,+}$ combines the block eigenvalues defined in Appendix~\ref{app:xy}. For diagonalizable $M$, write $M=V\operatorname{diag}(\mu_k)V^{-1}$ and $F=V^{-1}C(0)V^{-T}$. A Hermitian quadratic observable $O=O_0I+(\ii/4)w^TBw$, with real $O_0$ and a real antisymmetric matrix $B$, obeys

\begin{align}
\langle O(t)\rangle&=O_0+\sum_{k<\ell}W_{k\ell}e^{\lambda_{k\ell}^{(2)}t},\label{eq:pair}\\
\lambda_{k\ell}^{(2)}&=\mu_k+\mu_\ell,\label{eq:pairlambda}\\
W_{k\ell}&=-\tfrac14F_{k\ell}\Tr\!\left[B(v_kv_{\ell}^T-v_{\ell}v_k^T)\right].\label{eq:weights}
\end{align}
Here $v_k$ is a column of $V$. The transpose, rather than the Hermitian conjugate, is required. We use $V^{-T}=(V^{-1})^T$, and $T^+=T^{P=+1}$. The symbol $\simeq$ in Eq.~\eqref{eq:similarity} denotes matrix similarity. These $\binom{2N}{2}$ pair eigenvalues describe the bilinear subspace, not the full even sector. The identity is separate. The odd-sector scalar shift is essential in Eq.~\eqref{eq:spectrum} but does not enter these even observables.

For each distinct pair sum, the weights $W_{k\ell}$ must first be added over all pairs with that eigenvalue. A nonzero purely imaginary pair sum contributes an oscillation only if this total weight is nonzero; zero pair sums give stationary contributions. Pair sums with negative real parts give the decaying remainder.

To choose the large-system plotting windows, we separate the non-decaying contribution from the transient $R_O(t)$. Equation~\eqref{eq:pair} gives the observable-dependent estimate

\begin{equation}
|R_O(t)|\le\sum_{\substack{k<\ell\\\operatorname{Re}\lambda_{k\ell}^{(2)}<0}}|W_{k\ell}|e^{\operatorname{Re}\lambda_{k\ell}^{(2)}t}.
\label{eq:transient-envelope}
\end{equation}

The final insets for the large systems begin after this envelope falls below $10^{-4}$ in the normalized observable for both compared sizes. Isolation is identified by boundary connectivity, not a numerical cutoff on small decay rates. The logarithmic main panels display sampled dynamics over the relaxation interval. The linear insets resolve the surviving oscillations.

For diagonalizable $M$, antisymmetry of $F=V^{-1}C(0)V^{-T}$ gives

\begin{equation}
C(t)=\sum_{k<\ell}F_{k\ell}e^{(\mu_k+\mu_\ell)t}(v_kv_{\ell}^T-v_{\ell}v_k^T).\label{eq:Cexpand}
\end{equation}

Contracting with $B$ yields Eqs.~\eqref{eq:pair} and \eqref{eq:weights}. For $B_{\rm all}$, the independent entries are $B_{2j,2j+1}=-2/(N-1)$.

To establish Eq.~\eqref{eq:similarity}, let the upper hopping of a physical component be $2u_j$ and that of $T^+$ be $t_j$, with $t_j^2=u_j^2$. A diagonal similarity $S^{-1}MS$, with $S=\operatorname{diag}(s_j)$ and successive phase ratio $s_{j+1}/s_j=\ii t_j/u_j$ maps the upper entry to $2\ii t_j$ and the lower entry to the same value. The diagonal entries already obey $-2\gamma=2\ii(\ii\gamma)$. If a bond vanishes, this construction applies separately on each remaining connected segment. When $M$ is not diagonalizable, the matrix-exponential covariance solution remains valid even when Eq.~\eqref{eq:Cexpand} must be replaced by a Jordan expansion~\cite{Prosen2010,Barthel2022}.

The distinction between a stationary contribution and oscillation can be seen without numerical diagonalization. For XY at $N=4$, define $\widehat B_{\rm odd}=(\sigma_1^x\sigma_2^x+\sigma_3^x\sigma_4^x)/2$. In the initially populated sector $Q_x=+1$, where $Q_x=\prod_j\sigma_j^x$, set $Z_{\rm eff}=\sigma_1^x\sigma_2^x=\sigma_3^x\sigma_4^x$ and $X_{\rm eff}=\sigma_2^y\sigma_3^y$. The Hamiltonian of the isolated path is $-2J_xZ_{\rm eff}-J_yX_{\rm eff}$, the all-$x$ preparation has $Z_{\rm eff}=+1$, and $B_{\rm odd}(t)=\langle\widehat B_{\rm odd}\rangle_t$.

The effective two-level evolution gives the normalized signal,

\begin{equation}
f(t;u,v)=\frac{4u^2+v^2\cos(2\sqrt{4u^2+v^2}\,t)}{4u^2+v^2},\label{eq:minimal}
\end{equation}
with $(f,u,v)=(B_{\rm odd},J_x,J_y)$. The constant $4u^2/(4u^2+v^2)$ is a stationary contribution. The remaining term is a contribution at a nonzero frequency when $v\ne0$. Both are independent of $\gamma_L,\gamma_R$.

\subsection{Results for additional parameters}\label{app:regime}

\begin{figure}[t]

\centering

\includegraphics[width=0.8\columnwidth]{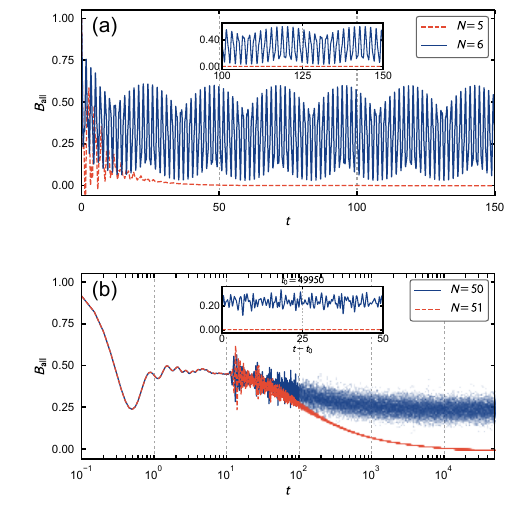}

\caption{XY spin correlation dynamics at $J_x=1$, $J_y=1.5$, and $\gamma_L=\gamma_R=0.2$. The initial state is $\ket{+x}^{\otimes N}$. (a) $N=5,6$. Inset: $100\le t\le150$. (b) $N=50,51$. Inset: $49950\le t\le50000$.}\label{fig:xy15}

\end{figure}

Figure~\ref{fig:xy15} gives the corresponding $J_y=1.5$ comparison. As in the main text, the even chain supports the non-decaying bond correlation. Changing the coupling changes its waveform but not the connectivity responsible for the non-decaying contribution.

For the large chains at this coupling ratio, the time window ends at $5\times10^4$, and the transient envelope in each final inset is below $10^{-4}$.

\section{Cluster chain}\label{app:cluster}

The general construction in Appendix~\ref{app:xy} also applies to the $r=2$ cluster chain with the same endpoint channels. We give only the model-dependent steps below.

\subsection{Majorana representation and odd-even effects}\label{app:cluster-paths}
For $r=2$, the physical Hamiltonian couples $a_j$ to $b_j$ through the field and $b_j$ to $a_{j+2}$ through the cluster interaction. The Majorana graph therefore separates into one path containing both Majoranas on every odd-indexed site and another containing both Majoranas on every even-indexed site. In the even operator sector, the boundary dissipators act through $a_1$ and $b_N$, exactly as in the XY chain. The Majorana operator $a_1$ always belongs to the odd-site path. For odd $N$, $b_N$ belongs to that same path, leaving the entire even-site path isolated; for even $N$, it belongs to the even-site path, so both paths are subject to boundary dissipation.

This explains the reversal relative to the XY chain. The XY paths alternate Majorana species between adjacent sites: $a_1$ and $b_N$ lie on the same path for even $N$. The cluster paths instead connect sites separated by two lattice spacings, so these two dissipative Majorana coordinates lie on the same path for odd $N$. Applying the criterion for undamped modes of Appendix~\ref{app:contact} therefore gives an isolated path for even XY chains and odd cluster chains, despite identical endpoint jump operators. For the latter, Eq.~\eqref{eq:exactright} ensures that even products of Majoranas on the isolated even-site path remain unaffected by dissipation. The odd-even reversal thus follows from the different bulk connectivity and requires no change in the boundary channels.

\subsection{Liouvillian spectrum}

For the physical Majoranas, $\sigma_j^z=-\ii a_jb_j$ and $\sigma_j^x\sigma_{j+1}^z\sigma_{j+2}^x=-\ii b_ja_{j+2}$. Using the vectorization and auxiliary fermions of Appendix~\ref{app:xy-spectrum} gives

\begin{equation}
\begin{aligned}
\LL_{{\rm cl},P}={}&2\ii h\sum_{j=1}^N(c_{2j-1}^\dagger c_{2j}+\mathrm{H.c.})\\
&-2\ii J\sum_{j=1}^{N-2}(c_{2j}^\dagger c_{2j+3}+\mathrm{H.c.})\\
&-2P\gamma_Ln_1-2\gamma_Rn_{2N}+\gamma_L(P-1).
\end{aligned}\label{eq:auxcl}
\end{equation}

The intersite hopping initially has sign $-J$. Set $g_j=(-1)^{\lfloor(j-1)/2\rfloor}$ and $\widetilde c_{2j-1}=g_jc_{2j-1}$, $\widetilde c_{2j}=g_jc_{2j}$. Since $g_{j+2}=-g_j$, this changes $-J$ to $J$ while leaving $h$, the occupations, the endpoint terms, and the global operator parity unchanged.

Reorder by $f_{o,2\ell-1}=\widetilde c_{4\ell-3}$, $f_{o,2\ell}=\widetilde c_{4\ell-2}$ and $f_{e,2\ell-1}=\widetilde c_{4\ell-1}$, $f_{e,2\ell}=\widetilde c_{4\ell}$. Each physical site contributes one $h$ bond. Set $\delta_N=[1-(-1)^N]/2$, $d_o=N+\delta_N$, and $d_e=N-\delta_N$. The reordered blocks are

\begin{align}
T_o^P&=S_{d_o}(h,J)+\ii P\gamma_LE_1^{(d_o)}+\ii\delta_N\gamma_RE_{d_o}^{(d_o)},\nonumber\\
T_e^P&=S_{d_e}(h,J)+\ii(1-\delta_N)\gamma_RE_{d_e}^{(d_e)}.\label{eq:CLblocks}
\end{align}

The operator-parity projection and spectral reconstruction derived in Appendix~\ref{app:xy-spectrum} are unchanged, so Eq.~\eqref{eq:spectrum} applies directly to these blocks. Their boundary terms encode the distribution of boundary dissipation between the paths established in Appendix~\ref{app:cluster-paths}: for odd $N$, $T_e^P$ has no dissipative boundary potential; for even $N$, each block has one.

\subsection{Magnetization dynamics}\label{app:cluster-dynamics}
The covariance derivation and spectral expansion in Appendix~\ref{app:covariance} apply without change. Only the Hamiltonian entries, initial covariance, and measured observable differ:
\begin{align}
A_{2j-1,2j}&=2h,\qquad A_{2j,2j+3}=2J,\nonumber\\
C_z(0)&=\bigoplus_{j=1}^N(-\mathsf J).
\end{align}
The remaining entries of $A$ follow by antisymmetry. The initial state is $\ket{\uparrow_z}^{\otimes N}$, and $\langle\sigma_j^z\rangle=-C_{2j-1,2j}$. For the mean magnetization, the observable matrix has $B_{2j-1,2j}=-2/N$. Substituting these quantities into Eqs.~\eqref{eq:pair}--\eqref{eq:weights} gives the magnetization response; Eq.~\eqref{eq:transient-envelope} bounds its decaying remainder.

For the cluster chain at $N=5$, restrict to the initially populated sector $\sigma_2^z\sigma_4^z=+1$ and set $Z_{\rm eff}=\sigma_2^z=\sigma_4^z$ and $X_{\rm eff}=\sigma_2^x\sigma_3^z\sigma_4^x$. The Hamiltonian of the isolated path is $-2hZ_{\rm eff}-JX_{\rm eff}$, and the all-$z$ preparation has $Z_{\rm eff}=+1$. The measured even-sublattice magnetization is $\langle Z_{\rm eff}\rangle_t$. These effective Pauli operators are representations within the indicated sectors.

The effective Hamiltonian has the same form as in the XY derivation, so Eq.~\eqref{eq:minimal} gives the result with $(f,u,v)=(m_{\rm even},h,J)$.

\subsection{Results for additional parameters}\label{app:cluster-regime}

\begin{figure}[t]

\centering

\includegraphics[width=0.8\columnwidth]{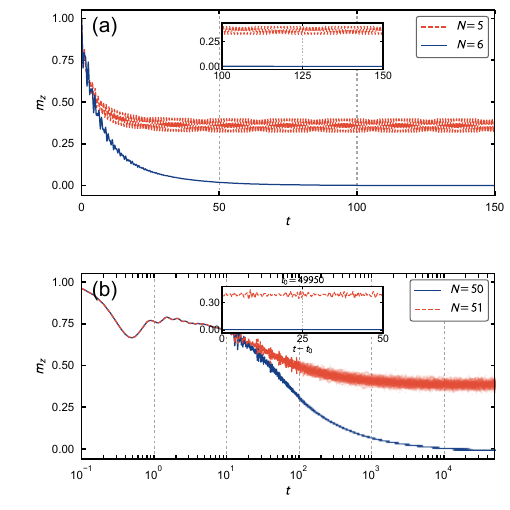}

\caption{Cluster magnetization at $J=1$, $h=1.5$, and $\gamma_L=\gamma_R=0.2$. The initial state is $\ket{\uparrow_z}^{\otimes N}$. (a) $N=5,6$. Inset: $100\le t\le150$. (b) $N=50,51$. Inset: $49950\le t\le50000$.}\label{fig:cluster15}

\end{figure}

Figure~\ref{fig:cluster15} repeats the size comparison of Fig.~\ref{fig:cluster} at $h=1.5$. The isolated-path assignment is unchanged: odd chains retain a non-decaying contribution, while even-chain magnetization relaxes. The changed coupling modifies the mean and oscillation amplitude while preserving the isolation from boundary dissipation.

For the large chains at this coupling ratio, the time window ends at $5\times10^4$, and the transient envelope in each final inset is below $10^{-4}$.

\subsection{Unequal boundary dissipation rates}\label{app:unequal}

The mechanism of isolation from boundary dissipation requires positive endpoint dissipation rates, but not equality between them. We illustrate this with the cluster chain with $r=2$ at $J=1$, $h=0.5$, $\gamma_L=0.2$, and $\gamma_R=0.1$. Changing the rates leaves the Majorana coordinates subject to boundary dissipation unchanged and therefore preserves which paths are isolated. For odd $N$, the isolated path and its Hamiltonian are unchanged, so its persistent oscillation frequencies and non-decaying magnetization contribution coincide with those at equal rates. The paths subject to boundary dissipation acquire different decay rates.

\begin{figure}[t]

\centering

\includegraphics[width=\columnwidth]{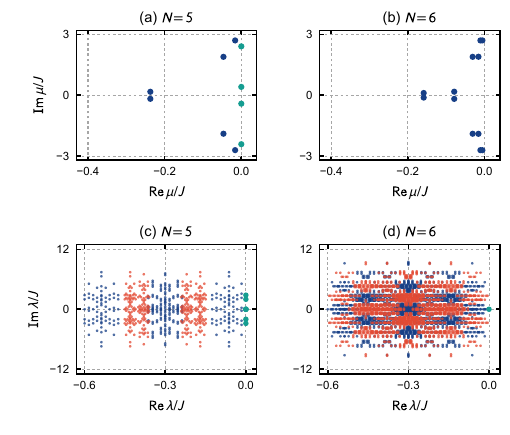}

\caption{Cluster spectra at $J=1$, $h=0.5$, $\gamma_L=0.2$, and $\gamma_R=0.1$. Columns: $N=5,6$. (a,b) Rapidity spectra. (c,d) Liouvillian spectra.}\label{fig:unequal-spectrum}

\end{figure}

Figures~\ref{fig:unequal-spectrum}(a,b) show the rapidity spectra for $N=5,6$. The odd chain retains four purely imaginary rapidities, consistent with $d_{\rm cl}=N-1$, whereas every rapidity of the even chain has negative real part. Figures~\ref{fig:unequal-spectrum}(c,d) show the corresponding full Liouvillian spectra. The odd chain supports nonzero purely imaginary eigenvalues alongside stationary modes at the origin, whereas only stationary modes remain non-decaying in the even chain. Thus unequal positive rates preserve the same spectral distinction as equal rates.

\begin{figure}[t]

\centering

\includegraphics[width=0.8\columnwidth]{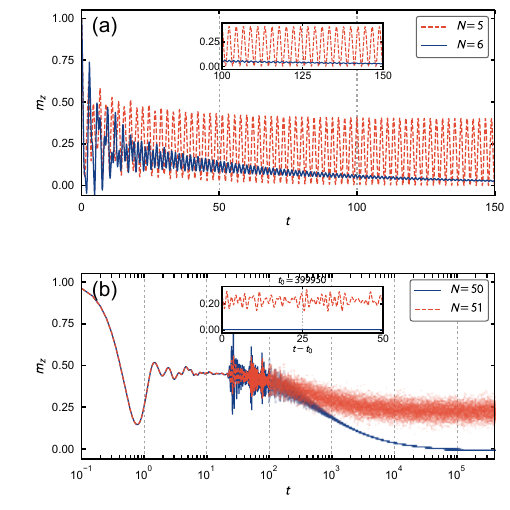}

\caption{Cluster magnetization at the unequal boundary dissipation rates of Fig.~\ref{fig:unequal-spectrum}. The initial state is $\ket{\uparrow_z}^{\otimes N}$. (a) $N=5,6$. Inset: $100\le t\le150$. (b) $N=50,51$. Inset: $399950\le t\le400000$.}\label{fig:unequal-dynamics}

\end{figure}

Figure~\ref{fig:unequal-dynamics} gives the corresponding dynamics from the all-$z$ product state for both small and large chains. The $N=6$ signal still contains a visible decaying contribution in the short-time inset. For $N=50,51$, the logarithmic time window extends to $4\times10^5$. The transient envelope in the final inset is below $4.8\times10^{-7}$ for both sizes. The odd-chain oscillations survive because the isolated paths remain unaffected by the unequal boundary dissipation rates.

\section{Cluster chains with even interaction ranges}\label{app:family}

\subsection{Model and boundary dissipation}

\begin{figure}[t]

\centering

\includegraphics[width=\columnwidth]{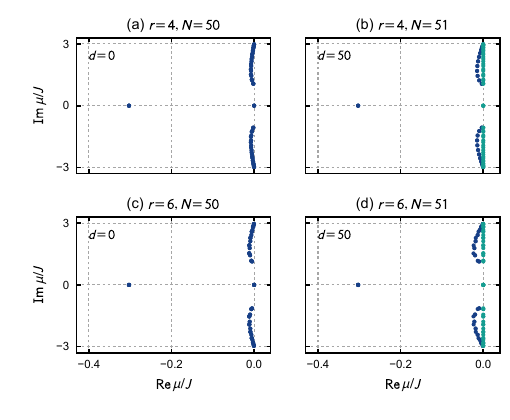}

\caption{Cluster rapidities for even interaction ranges at $J=1$, $h=0.5$, and $\gamma_{L,k}=\gamma_{R,k}=0.2$. Rows: $r=4,6$. Columns: $N=50,51$. Blue/green denote damped/undamped modes. Here $d$ is the number of undamped modes, including degeneracies.}\label{fig:family-rapidities}

\end{figure}

For Eq.~\eqref{eq:Hr-family}, the physical Jordan--Wigner transformation gives

\begin{equation}
H_r=\ii h\sum_{j=1}^N a_jb_j+\ii J\sum_{j=1}^{N-r}b_ja_{j+r}.
\label{eq:family-majorana}
\end{equation}

The path labeled $q=1,\ldots,r$ contains sites $q,q+r,q+2r,\ldots$ up to $N$, with alternating on-site $h$ and intersite $J$ links. For even $r$, each path contains only odd-indexed or only even-indexed physical sites, which underlies the choice of boundary jump operators used here. For odd $r$, successive sites on a path alternate between odd and even indices, and the endpoint assignment depends on the residue classes modulo $r$. The same construction therefore does not directly yield the odd-even dependence established above. We restrict the extension of the cluster family to even $r$ and leave the classification of boundary channels for odd ranges to future work.

The jumps of Eq.~\eqref{eq:family-jumps} obey

\begin{align}
L_{L,k}/\sqrt{\gamma_{L,k}}&=a_{2k+1},
L_{R,k}/\sqrt{\gamma_{R,k}}&=\ii P_s b_{N-2k}.
\label{eq:family-jw-jumps}
\end{align}
The right jump can therefore be replaced by the associated $b$ Majorana only when evaluating the dissipative action on operators with even operator parity. The rapidity spectrum matrix below describes the evolution of even bilinears. Even products of its isolated Majorana coordinates generate non-decaying observables. The covariance equation retains the form of Eq.~\eqref{eq:covdyn}, with

\begin{equation}
\begin{aligned}
M_r&=A_r-2D_r,\\
D_r&=\sum_{k=0}^{r/2-1}\bigl(\gamma_{L,k}e_{a_{2k+1}}e_{a_{2k+1}}^T\\
&\hspace{27mm}+\gamma_{R,k}e_{b_{N-2k}}e_{b_{N-2k}}^T\bigr).
\end{aligned}\label{eq:family-drift}
\end{equation}
Here $e_{a_j},e_{b_j}$ select the corresponding physical Majorana coordinates, and $A_r$ is defined by $H_r=(\ii/4)w^TA_rw$. In particular, $(A_r)_{a_jb_j}=2h$ and $(A_r)_{b_ja_{j+r}}=2J$, with the opposite entries fixed by antisymmetry.

\subsection{Spectrum and dynamics}

\begin{figure}[t]

\centering

\includegraphics[width=0.8\columnwidth]{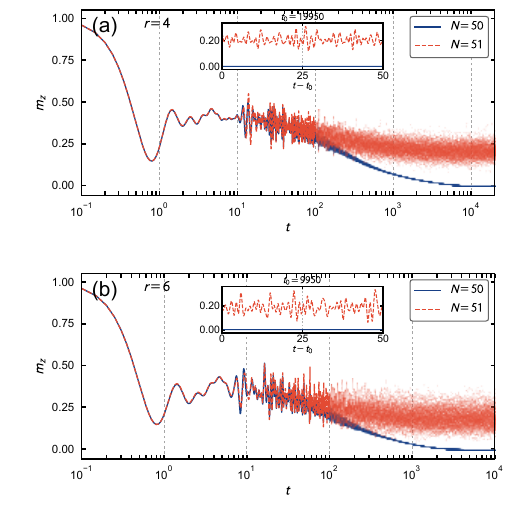}

\caption{Cluster magnetization for even interaction ranges from $\ket{\uparrow_z}^{\otimes N}$ at the parameters of Fig.~\ref{fig:family-rapidities}, comparing $N=50,51$. (a) $r=4$. Inset: $19950\le t\le20000$. (b) $r=6$. Inset: $9950\le t\le10000$.}\label{fig:family-dynamics}

\end{figure}

For odd $N$, both sets of boundary dissipators act on paths on odd sites. For even $N$, they act on complementary sets of paths. Equation~\eqref{eq:contact-decay} applies to $M_r$ with $D=D_r$. Since $Jh\ne0$, endpoint recursion excludes every path subject to boundary dissipation from the undamped subspace, yielding Eq.~\eqref{eq:family-d}.

For these jumps, the Heisenberg-picture evolution preserves the number of Majorana factors in an even product. For even $k$, the adjoint Liouvillian restricted to the subspace spanned by products of $k$ distinct Majorana operators has eigenvalues given by sums of $k$ distinct rapidities. Any sum involving a damped mode has negative real part. Thus the non-decaying operators with even operator parity are built entirely from isolated modes, together with the identity.

Figure~\ref{fig:family-rapidities} shows the rapidity spectra for $r=4,6$ at $N=50,51$. For each range, all rapidities of the even chain have negative real parts, whereas the odd chain has 50 purely imaginary rapidities, counting multiplicities. These counts agree with $d_{\rm cl}=0$ and $d_{\rm cl}=N-1$, respectively. Even products of the isolated modes evolve without dissipation and can contribute to the magnetization response.

Figure~\ref{fig:family-dynamics} supplies the corresponding dynamical evidence at identical parameters, using the all-$z$ initial state. As in the cluster chain with $r=2$, even-site magnetization accesses the isolated paths of odd chains. The large-system windows extend to $2\times10^4$ for $r=4$ and $10^4$ for $r=6$. In each final inset the absolute transient envelope is below $10^{-4}$ for both sizes, exposing the odd-chain oscillations after the damped contribution has become negligible.

\end{document}